\documentclass[submission,copyright,creativecommons,blind]{eptcs}
\providecommand{\event}{SOS 2007} 

\usepackage{iftex}

\ifpdf
  \usepackage{underscore}         
  \usepackage[T1]{fontenc}        
\else
  \usepackage{breakurl}           
\fi

\usepackage{amsmath}
\usepackage{amsthm}
\usepackage{amsfonts}
\usepackage{amssymb}
\usepackage{tikz}
\usepackage[most]{tcolorbox}
\usepackage{graphicx}
\usepackage{bussproofs}
\usepackage{listings}
\usepackage{textgreek}
\usepackage{parcolumns}
\usepackage{float}
\usepackage{stackengine}
\usepackage{empheq}

\newcommand\code[1]{{\tt\small #1}}
\definecolor{dkgreen}{rgb}{0,0.6,0}
\definecolor{ltblue}{rgb}{0,0.4,0.4}
\definecolor{dkviolet}{rgb}{0.3,0,0.5}
\definecolor{dkblue}{rgb}{0,0,1}
\definecolor{dkred}{rgb}{1,0,0}

\lstdefinelanguage{Coq}{ 
    mathescape=true,
    texcl=false, 
    morekeywords=[1]{Section, Module, End, Require, Import, Export,
        Variable, Variables, Parameter, Parameters, Axiom, Hypothesis,
        Hypotheses, Notation, Local, Tactic, Reserved, Scope, Open, Close,
        Bind, Delimit, Definition, Let, Ltac, Fixpoint, CoFixpoint, Add,
        Morphism, Relation, Implicit, Arguments, Unset, Contextual,
        Strict, Prenex, Implicits, Inductive, CoInductive, Record,
        Structure, Canonical, Coercion, Context, Class, Global, Instance,
        Program, Infix, Theorem, Lemma, Corollary, Proposition, Fact,
        Remark, Example, Proof, Goal, Save, Qed, Defined, Hint, Resolve,
        Rewrite, View, Search, Show, Print, Printing, All, Eval, Check,
        Projections, inside, outside, Def},
    morekeywords=[2]{forall, exists, exists2, fun, fix, cofix, struct,
        match, with, end, as, in, return, let, if, is, then, else, for, of,
        nosimpl, when},
    morekeywords=[3]{Type, Prop, Set, true, false, option},
    morekeywords=[4]{pose, set, move, case, elim, apply, clear, hnf,
        intro, intros, generalize, rename, pattern, after, destruct,
        induction, using, refine, inversion, injection, rewrite, congr,
        unlock, compute, ring, field, fourier, replace, fold, unfold,
        change, cutrewrite, simpl, have, suff, wlog, suffices, without,
        loss, nat_norm, assert, cut, trivial, revert, bool_congr, nat_congr,
        symmetry, transitivity, auto, split, left, right, autorewrite},
    morekeywords=[5]{by, done, exact, reflexivity, tauto, romega, omega,
        assumption, solve, contradiction, discriminate},
    morekeywords=[6]{do, last, first, try, idtac, repeat},
    morecomment=[s]{(*}{*)},
    showstringspaces=false,
    morestring=[b]",
    morestring=[d]’,
    tabsize=3,
    extendedchars=false,
    sensitive=true,
    breaklines=false,
    basicstyle=\small,
    captionpos=b,
    columns=[l]flexible,
    identifierstyle={\ttfamily\color{black}},
    keywordstyle=[1]{\ttfamily\color{dkviolet}},
    keywordstyle=[2]{\ttfamily\color{dkgreen}},
    keywordstyle=[3]{\ttfamily\color{ltblue}},
    keywordstyle=[4]{\ttfamily\color{dkblue}},
    keywordstyle=[5]{\ttfamily\color{dkred}},
    stringstyle=\ttfamily,
    commentstyle={\ttfamily\color{dkgreen}},
    literate=
    {\\forall}{{\color{dkgreen}{$\forall\;$}}}1
    {\\exists}{{$\exists\;$}}1
    {<-}{{$\leftarrow\;$}}1
    {=>}{{$\Rightarrow\;$}}1
    {==}{{\code{==}\;}}1
    {==>}{{\code{==>}\;}}1
    {->}{{$\rightarrow\;$}}1
    {<->}{{$\leftrightarrow\;$}}1
    {<==}{{$\leq\;$}}1
    {\#}{{$^\star$}}1 
    {\\o}{{$\circ\;$}}1 
    {\@}{{$\cdot$}}1 
    {\/\\}{{$\wedge\;$}}1
    {\\\/}{{$\vee\;$}}1
    {++}{{\code{++}}}1
    {~}{{$\sim$}}1
    {\@\@}{{$@$}}1
    {\\mapsto}{{$\mapsto\;$}}1
    {\\hline}{{\rule{\linewidth}{0.5pt}}}1
}[keywords,comments,strings]

\DeclareFontFamily{U}{mathb}{\hyphenchar\font45}
\DeclareFontShape{U}{mathb}{m}{n}{
<-6> mathb5 <6-7> mathb6 <7-8> mathb7
<8-9> mathb8 <9-10> mathb9
<10-12> mathb10 <12-> mathb12
}{}
\DeclareSymbolFont{mathb}{U}{mathb}{m}{n}
\DeclareMathSymbol{\ggcurly}{\mathrel}{mathb}{"CF}

\usepackage[noend]{algorithmic}
\usepackage{subcaption}
\usepackage{hyperref}
\usepackage[export]{adjustbox}

\newcommand\sigmamu[0]{\sigma_{\texttt{min}}}
\newcommand\mumatch[2]{\sigmamu({#1}, \ {#2})}

\newcommand\grafting[2]{{#1} \ {#2}}

\newcommand\asapply{\texttt{as\textunderscore apply}}

\theoremstyle{definition}
\newtheorem{example}{Example}

\title{Work-in-Progress: A Tactic for Pattern Matching in Autosubst}
\author{Mathews George
	\institute{Heriot-Watt University\\
		Edinburgh, UK}
	\email{mg2079@hw.ac.uk}
	\and
	 Kathrin Stark
	\institute{Heriot-Watt University\\
		Edinburgh, UK}
	\email{\quad k.stark@hw.ac.uk}
}
\def\titlerunning{A Tactic for Pattern Matching in Autosubst}
\def\authorrunning{M. George et al.}

\begin{document}

\maketitle

\begin{abstract}
    Autosubst enables automatic equality-checking up to the $\sigma$-calculus for assumption-free equalities, allowing users to avoid cumbersome reasoning about de Bruijn indices. While effective in many cases, this approach is inapplicable when matching against typing rules, reduction relations, or lemmas, requiring users to either phrase typing rules in a way that they work with Autosubst or even stating explicitly an alternative de Bruijn term. 
    But even without $\beta$-reduction, solutions of matching may not be unique.
    This paper presents a work-in-progress method for automatically pattern matching against assumptions, evaluated on standard case studies including the POPLMark and POPLMark Reloaded challenges.
\end{abstract}

\section{Introduction} \label{intro}

De Bruijn terms are a canonical technique for representing terms with binders~\cite{de1972lambda}, 
and have a long history of being used in mechanised formalisations of languages with binders in proof assistants. 
Despite this, they are often as regarded as cumbersome in practice
\cite{aydemir2005mechanized}, 
largely due to the overhead involved in proving and applying the many technical lemmas required for substitution and renaming. 

The Autosubst code generator \cite{schafer2015autosubst,stark2019autosubst}  addresses this issue by automating reasoning about de Bruijn terms in the Rocq (formerly Coq) proof assistant in many cases. 
Given a signature describing a custom syntax, Autosubst automatically proves the substitution lemmas corresponding to the equations in the $\sigma$-calculus \cite{abadi1989explicit}. 
Central to the approach, Autosubst comes with a simplification tactic called \texttt{asimpl} which normalizes terms according to these equations.
As the de Bruijn algebra forms a sound and complete model of the $\sigma$-calculus \cite{schafer2015}, and as the rewriting system is convergent \cite{curien1996confluence}, equality in the de Bruijn algebra is decidable by this tactic.
In summary, this allows users to work with de Bruijn syntax without manually applying many technical lemmas. 

However, this approach becomes less effective in situations where we want to unify a goal with an assumption, in Rocq's case via the \texttt{apply} or \texttt{eapply} tactic. 
In particular, Autosubst's automation is designed for assumption-free reasoning and does not support matching modulo $\sigma$-equivalence. 
As a consequence, the user must compensate for these limitations in their proof scripts. Common workarounds include stating rules in an indirect form to facilitate matching, introducing auxiliary lemmas with the right instantiations, and manually transforming goals during proofs to make them applicable, together with directly defining the values of appearing existential variables. 


A representative example, is the formulation of the rule for type application in System $F_{<:}$ in the POPLMark solution based on Autosubst:
	\begin{lstlisting}[language=Coq, escapeinside={(*}{*)}]
	(*$T\_Tapp$*) : (*$\forall \ \Delta \ \Gamma \ s \ A \ A' \ B \ B'$*), (*$\Delta; \Gamma \vdash s: \forall A. B \to$*) (*$\Delta \vdash A' <: A \to$*) (*$B' = B[A' \cdot id]$*) 
	            (* $ \to \Delta; \Gamma \vdash s \ A' : B'$*)
\end{lstlisting}
Here, the result type is expressed using a fresh meta-variable $B'$ rather than the more direct expression $B[A' \cdot id]$.
While this formulation enables the use of the \texttt{eapply} tactic, it shifts the burden to the user, who (when applying this rule to a goal) must subsequently discharge the additional equality premise.
This premise typically contains existential variables, 
and the \texttt{asimpl} tactic is only helpful in the case that the user can give the value of these existential variables upfront.

For example, proving the context morphism lemma in the POPLMark challenge, in the corresponding case we get stuck with the following goal 
	\begin{lstlisting}[language=Coq, escapeinside={(*}{*)}]
		(*$\Delta; \Gamma \vdash s[\sigma;\tau] \ A'[\sigma] : B[A'[\sigma] \cdot \sigma]$*)
	\end{lstlisting}
where, when applying \texttt{T_Tapp}, 
we need to solve the equation $B[A[\sigma] \cdot \sigma] = ?B[A[\sigma] \cdot id]$ with a constraint on $?B$ as $\Delta;\Gamma \vdash s[\sigma] : \forall A[\sigma]. ?B$.

Such indirect formulations are pervasive.
They arise not only in typing rules, but also in reduction relations~\cite{forster2019call, abel2017poplmark}, weakening and renaming lemmas, and even in mechanisations of dependent type theory~\cite{adjedj2024martin}.
Even when rules are stated directly, auxiliary lemmas are often introduced solely to make them compatible with \texttt{apply}.

This raises a natural question: can we recover the level of automation provided by Autosubst while supporting matching modulo $\sigma$-equivalence in the presence of assumptions?

The problem is inherently difficult.
Moura et al. showed that Dowek et al.'s method \cite{dowek1995higher} does not decide second-order matching for the full $\lambda\sigma$-calculus \cite{de2005second}, but even without $\beta$, solutions are not necessarily unique.
In practice, this means that a matching problem may admit multiple candidates, not all of which satisfy the premises in the context.


\begin{example}[Matching up to $\sigma$-equivalence is not unique.]
The $\sigma$-matching problem we need to solve is constrained by the premises in which the unknown variables appear.
In the above example, we need to solve the equation $B[A[\sigma] \cdot \sigma] = ?B[A[\sigma] \cdot id]$ with a constraint on $?B$ as $\Delta;\Gamma \vdash s[\sigma] : \forall A[\sigma]. ?B$. We have at least two solutions for $?B$, $B[\Uparrow \ \sigma]$ and $B[A[\sigma \circ \uparrow] \cdot \sigma \circ \uparrow]$. 
The latter solution does not hold with the premise. 
\end{example}

As a consequence, any practical solution must rely on heuristics.
In the worst case, an adversarial example can always be constructed where a heuristic selects an incorrect solution, even when a correct one exists.


\paragraph{Contributions.}
We extend the Autosubst framework with a tactic for pattern-matching modulo $\sigma$-equivalence.
Our main contributions are as follows:
\begin{enumerate}
\item We introduce a heuristic tactic \asapply, intended as a replacement for the \texttt{apply/eapply} tactic in Autosubst-based developments.
\item We generalize this tactic to support the full input language of Autosubst.
\item We evaluate our approach on several case studies, including the entire POPLMark and POPLMark Reloaded Challenge and show that despite its theoretical limitations, first results suggest it performs well in practice.
\end{enumerate}


The development including all results is available online \cite{lfmtpRepo}.

\section{Background}
We consider de Bruijn terms for the $\lambda$-calculus \cite{de1972lambda}:
\begin{align*}
    s,t \in \mathbb{T} \ ::= \ n \ | \ s \ t \ |  \ \lambda. s \qquad (n \in \mathbb{N})
\end{align*}
where $s \ t$ denotes application, $\lambda. s$ abstraction, and $n$ are de Bruijn indices which serve as variables bound by abstraction. 

Parallel substitutions $\sigma,\tau$  are total functions mapping indices to terms. Intuitively, a substitution can be seen as an infinite sequence of terms. Sometimes, we need to deal with a special kind of substitutions, written $\xi,\zeta$,  that map indices to indices known as \textit{renamings}.

A \textit{de Bruijn algebra} is a first-order two-sorted algebra formed of de Bruijn terms and substitutions, and certain constants and operations from the $\sigma$-calculus of Abadi et al.\,\cite{abadi1989explicit}.
The central operation is instantiation: 
 $s[\sigma]$ replaces the free indices of the term $s$ with the terms provided by the substitution $\sigma$. 

The $\sigma$-calculus comes with a defined set of primitives.
The cons operation $s \cdot \sigma$ prepends a term $s$ to a substitution $\sigma$. The composition of substitutions is defined as $ (\sigma \circ \tau) \ n \ := (\sigma \ n)[\tau]$. We further have constant substitutions such as the identity $id \ := \lambda n. n$ and the shift $\uparrow \ := \lambda n. (n+1)$. The lifting of a substitution is defined as and used as notation for $\Uparrow \ \sigma := 0 \cdot \sigma \circ \uparrow$.

The $\sigma$-calculus is equipped with a directed set of equational rules over these primitives (Figure \ref{fig:eqtheory}). We denote $s\equiv_{\sigma'}t$ if $s$ and $t$ are provably equal by rewriting with a fragment $\sigma'$ of $\sigma$-rules. We denote by $\sigmamu$ the fragment consisting of the rules inside the box in Figure~\ref{fig:eqtheory}, which characterise instantiation and its compositionality properties.

Sch\"afer et al. showed that the de Bruijn algebra forms a sound and complete model of the $\sigma$-calculus \cite{schafer2015}, and, as the rewriting system induced by these rules is known to be convergent \cite{curien1996confluence},  equality in the de Bruijn algebra is decidable by normalisation with respect to the $\sigma$-rules. 

\begin{figure}[b!]
	\begin{equation*}
        \boxed{\begin{aligned}[c]
			(s \ t) [\sigma] &\equiv s[\sigma] \ t[\sigma] \\
			(\lambda.s)[\sigma]&\equiv \lambda.(s[\Uparrow \sigma]) \\
			s[\sigma][\tau]&\equiv s[\sigma \circ \tau] \\	
			(s \cdot \sigma) \circ \tau &\equiv \ s[\tau] \cdot \sigma \circ  \tau \\
			(\sigma \circ \tau) \circ \rho &\equiv \sigma \circ \tau \circ \rho \\
		\end{aligned}}
		\ \ \ \quad
        \begin{aligned}[c]
			s[id] &\equiv s \\
			\sigma \circ id &\equiv \sigma \\
			id \circ \sigma &\equiv \sigma \\
            0 \ \cdot \uparrow &\equiv id
		\end{aligned}
        \ \ \ \quad
		\begin{aligned}[c]
           	0[s \cdot \sigma] &\equiv s \\
            \uparrow \circ \ (s \cdot \sigma) &\equiv \sigma \\
			0    [\sigma] \ \cdot \uparrow \circ \ \sigma &\equiv \sigma
		\end{aligned}
	\end{equation*}
	\caption{The rewriting system of the $\sigma$-calculus}\label{fig:eqtheory}
\end{figure}

We also require notions from $\sigma$-matching \cite{dowek1995higher, baader1998term}.
A $\sigma$-substitution $\theta$ is a function from the object variables to $\sigma$-expressions such that $\theta \ x \neq x$ for finitely many $x$. Its domain is  $\mathcal{D}(\theta) := \{ x \ | \ \theta \ x \neq x\}$. We write  $\theta$ as $\{x_1 \mapsto s_1,...,x_n \mapsto s_n\}$ where $x_i \in \mathcal{D}(\theta)$ and denote its application to a $\sigma$-expression $s$ simply as $\grafting{\theta}{s}$. Two substitutions $\theta_1$ and $\theta_2$ are $\sigma'$-equal iff  $\theta_1 \ x \equiv_{\sigma'}  \theta_2 \ x$ for all $x$.  Otherwise, $\theta_1$ and $\theta_2$ are $\sigma'$-different.  

We distinguish between \textit{bound} and \textit{free} object variables. 
The free variables  $\mathcal{F}(s)$ are the ones we need to resolve in a unification problem \cite{dowek1995higher}.
We use the notation $s=_{\sigma'} s'$ to denote a matching equation in a fragment $\sigma'$ of the $\sigma$-calculus where $s$ does not have free variables. We say $s=_{\sigma'}t$ has a solution $\theta$ iff $s \equiv_{\sigma'} \grafting{\theta}{t}$ and $\mathcal{D}(\theta) \subseteq \mathcal{F}(t)$. We call a $\sigma'$-matching algorithm $\sigma$-complete iff it computes all $\sigma$-different solutions to a $\sigma'$-matching problem.
In our intended setting, bound variables correspond to abstract meta-variables, while free variables correspond to unresolved existential variables in a matching problem.

\section{A Pattern Matching Tactic for Autosubst}

We describe a general pattern-matching tactic \asapply\, for terms of the $\lambda$-calculus.
The tactic is implemented in Rocq's tactic language Ltac.


Let $P$ be an $n$-ary relation, and consider the following Rocq goal:
\begin{tcolorbox}[colback=white!5!white,colframe=black!20!black]
\begin{lstlisting}[language=Coq, escapeinside={(*}{*)}]
(*$H \ : \ \forall $*) <qvars>, <premises> (*$ \to P \ t_1 \ t_2 \ ... \ t_n $*) 
-------------------------------------------
(*$P \ s_1 \ s_2 \ ... \ s_n$*)
\end{lstlisting}
\end{tcolorbox}
with $H$ an assumption in the context we want to apply, where \texttt{<qvars>} and \texttt{<premises>} denote the quantified variables and premises of $H$.

A call of \asapply \, $H$ proceeds in two phases: preprocessing and matching.

\paragraph{Preprocessing.}
We first normalise both the goal and the conclusion of $H$ with respect to the $\sigma$-calculus using the \texttt{asimpl} tactic provided by Autosubst.

Next, all quantified variables are replaced with existential variables (evars) of the corresponding types, which will be instantiated during matching. 
The premises of $H$ are then turned into subgoals, to be solved after matching with instantiated quantified variables. 

The actual matching will be between the corresponding $s_i$ and $t_i$. In a first step, we try whether this can be solved by pure syntactic matching using Rocq's tactic. 
Furthermore, if $s_i$ uses Autosubst's explicit renamings (i.e., is of the form 
$s\langle\xi\rangle$ or $s[\xi]$) but $t_i$ is not, we change 
$s_i$ to $s[\xi]$ ($s\langle\xi\rangle$) using \texttt{substify} (\texttt{renamify}). 
These are Autosubst-provided tactics that perform conversion between instantiation and renaming operations.



\paragraph{Matching Phase.}

After preprocessing, we obtain a  $\sigma$-matching problem 
\[\mathcal{M}=\{s_i=_\sigma t_i \ | \ s_i \textnormal{ and } t_i  \textnormal{ are } \sigma\textnormal{-expressions} \}\]
We process equations in $\mathcal{M}$ from left to right.

For each equation $s_i =_\sigma t_i$, we first try to match in the $\sigmamu$ fragment using a procedure $\mumatch{s_i}{t_i}$. If successful, this yields a substitution $\theta$, which is applied to all remaining equations $t_k$ for $k \geq i$.



The function $\mumatch{s}{t}$  is a partial procedure that computes substitutions by matching modulo the $\sigmamu$ fragment. It proceeds as follows:
\begin{itemize}
\item If $s$ and $t$ match syntactically, a substitution is returned.
\item Otherwise, $t$ is rewritten using $\sigmamu$ rules (Figure~\ref{fig:eqtheory}), and matching is retried.
\end{itemize}


In Figure \ref{fig:musig}, we give an operational description of the $\sigmamu$-function. The function performs a backtracking search over possible rewrites of $t$ until a match with $s$ is found.
For example, $\sigmamu(s[\sigma\circ \tau], \ ?s[\tau])$ matches first with the seventh case, but the execution path stemming from it does not result in a $\sigma$-substitution. The execution backtracks and continues by matching with the tenth case, resulting in $\{?s \mapsto s[\sigma]\}$.

If $\sigmamu(s,t)$ fails, we apply a small set of heuristics capturing common patterns in developments:
\begin{itemize}
\item If $s_i =_\sigma t_i$ is of the form $s[t \cdot \sigma] =_\sigma ?s[t \cdot id]$, instantiate $t_k$ with $\{?s \mapsto s[\Uparrow \ \sigma]\}$, for $k \geq i$.
\item If $s_i =_\sigma t_i$ is of the form $s =_\sigma ?s[?\sigma]$, instantiate $t_k$ with $\{?s \mapsto s, \ \sigma \mapsto id\}$ respectively, for $k \geq i$.
\item If $s_i =_\sigma t_i$ is of the form $s =_\sigma ?s[id]$, instantiate $t_k$ with $\{?s \mapsto s\}$ respectively, for $k \geq i$.
\item If $s_i =_\sigma t_i$ is of the form $s =_\sigma s[?\sigma]$, instantiate $t_k$ with $\{?\sigma \mapsto id\}$ respectively, for $k \geq i$.
\end{itemize}

 
We again $\sigma$-normalize $H$ with the potentially instantiated existential variables. The tactic succeeds if the goal and $H$ are syntactically equal at this point.

 \begin{figure*}
	\begin{equation*} 
		\begin{aligned}[c]
          \mumatch{s}{s} \ &\Rightarrow \ \{\} \\
		   \mumatch{s}{?s} \ &\Rightarrow \ \{?s \mapsto s\} \\
            \mumatch{\lambda.s}{\lambda. s'} \ &\Rightarrow \ \mumatch{s}{s'} \\
	          \mumatch{s \ t} {s' \ t'} \ &\Rightarrow \ \mumatch{t}{\grafting{\mumatch{s}{s'}}{t'}} \  \cup \  \mumatch{s}{s'}\\
            \mumatch{\sigma \circ \tau} {\sigma' \circ \tau'} \ &\Rightarrow  \ \mumatch{\tau} {\grafting{\mumatch{\sigma} {\sigma'}}{\tau'}} \ \cup \ \mumatch{\sigma}{\sigma'} \\
             \mumatch{s \cdot \sigma}{t \cdot \tau}  \ &\Rightarrow \ \mumatch{\sigma}{\grafting{\mumatch{s}{t}}{\tau}} \ \cup  \ \mumatch{s}{t} \\ 
            \mumatch{s[\sigma]}{s'[\sigma']} \ &\Rightarrow \ \mumatch{\sigma}{\grafting{\mumatch{s}{s'}}{\sigma'}} \ \cup \ \mumatch{s}{s'} \\
        \mumatch{\lambda.s[\Uparrow \ \sigma]}{s' [\sigma']} \ &\Rightarrow \ \mumatch{(\lambda.s)[\sigma]} {s'[\sigma']} \\
        \mumatch{s[\sigma] \ t[\sigma]}{s'[\sigma]} \ &\Rightarrow \ \mumatch{(s \ t)[\sigma]}{s'[\sigma]}\\
          \mumatch{s[\sigma \circ \tau]} {s'[\sigma']} \ &\Rightarrow \ \mumatch{s[\sigma][\tau]}{s'[\sigma]} \\
        \mumatch{\sigma\circ \tau\circ \rho}{\sigma' \circ \tau'} \ &\Rightarrow  \ \mumatch{(\sigma \circ \tau) \circ \rho} {\sigma' \circ \tau'} \\
          \mumatch{s[\tau] \cdot \sigma \circ \tau} {\sigma'\circ\tau'} \ &\Rightarrow \ \mumatch{(s \cdot \sigma) \circ \tau} {\sigma'\circ \tau'}
		\end{aligned}
		\ \ \ \quad
	\end{equation*}
	\caption{The $\sigmamu$-function for matching in the $\sigmamu$ fragment}\label{fig:musig}
\end{figure*}

\subsection{Generalisation} 

Autosubst \cite{stark2019autosubst} is a code generator that takes a second-order HOAS \cite{pfenning1988higher} representation of a custom syntax and generates a model of extended $\sigma$-calculus comprising multiple mutually inductive sorts, vector substitutions, and first-class renamings. 


This tactic has been generalised to work for the input language of the Autosubst compiler.
The $\sigmamu$ fragment describes the instantiation operation and has compositionality laws. Hence, for a general syntax, the $\sigmamu$-function has congruence cases, and cases corresponding to the instantiation and compositionality laws.  

If $\sigmamu(s,t)$ fails, we check whether $s=_\sigma t$ matches with certain equations that are recurring in developments. In the case of the first equation $s[t \cdot \sigma]=_\sigma ?s[t \cdot id]$, the expression $s[t \cdot id]$ denotes the elimination of the abstraction $\lambda.s$ by substituting $t$ for the bound index. We generate a case accordingly for each binder in the syntax. For example, if we have a polyadic binder $\lambda_2 : (tm \to \ tm \to tm) \to tm$, we generate a case for the equation $s[t_1 \cdot t_2 \cdot \sigma] = ?s[t_1 \cdot t_2 \cdot id]$. In the solution $\{?s \mapsto s[\Uparrow \ \sigma]\}$ for the $\lambda$-abstraction, the substitution $\Uparrow \sigma$ comes from the instantiation law $(\lambda.s)[\sigma]=\lambda.s[\Uparrow \ \sigma]$. For a general binder, this would be the lifted substitution vector formed during instantiation. In the case of $\lambda_2$, the solution would be $\{?s \mapsto s[\Uparrow\Uparrow \sigma]\}$ because $(\lambda_2.s)[\sigma] = \lambda_2.s[\Uparrow \Uparrow \ \sigma]$.

Note that as an additional difficulty since Autosubst supports first-class renamings, the $\sigmamu$-function and the specific equations have cases for the renaming operation, which we omit here for conciseness.

\subsection{Current Limitations}
We summarise some of the limitations of matching in the $\sigmamu$ fragment and the \asapply\ tactic.

 First, note that, per se, the $\sigmamu$ fragment is not a confluent rewrite system. The expression $(\lambda.s)[\sigma][\tau]$ reduces to both $\lambda.s[0[\Uparrow \ \tau] \cdot \sigma \circ \uparrow \circ \Uparrow \tau ]$ and $ \lambda.s[0 \cdot \sigma \circ \tau \circ \uparrow]$ via $\sigmamu$ rules. But, they are not joinable as $\sigmamu$ lacks the rules $0[s \cdot \sigma] \rightarrow s$ and $\uparrow \circ (s \cdot \sigma) \rightarrow \sigma$. Of course they are joinable via the whole $\sigma$-calculus. 
 
Note that $\sigmamu$-matching does not have unitary solutions in general. An example is $s[\sigma \circ \tau]=_{\sigmamu}?s[?\sigma]$ which has two solutions $\{?s \mapsto s, \ ?\sigma \mapsto \sigma \circ \tau\}$ and $\{?s \mapsto s[\sigma], \ ?\sigma \mapsto \tau \}$.

Generally,
\asapply\ is not $\sigma$-complete. In the above example, it would only produce the first solution.
\asapply\ is not complete even as a decider. It fails for the equation $t=_\sigma? s[t \cdot \sigma]$ though there is the solution $\{?s \mapsto 0\}$. Note that this would rarely be a problem in a practical development.

\section{Case Studies}


We have tested the tactic on Rocq solutions of the POPLMark \cite{aydemir2005mechanized} and POPLMark Reloaded \cite{abel2017poplmark} challenges, two benchmarks on reasoning with binders. In this section, we present examples from our modified solutions. 
The linked development contains both the original and new solutions.

\begin{example} When proving substitutivity of multi-step in the POPLMark Reloaded, we have to prove:
\begin{lstlisting}[language=Coq, escapeinside={(*}{*)}]
	(*$mstep\_inst$*) : (*$\forall \ s \ t \ \sigma$*), (*$s \ggcurly t \to$*) (*$s[\sigma] \ggcurly t[\sigma]$*)
	(*$H$*) : (*$\forall n$*), (*$\sigma \ n \ggcurly \tau \ n$*)
	--------------------------------------
	(*$(\sigma \ i)\langle\uparrow\rangle \ggcurly (\tau \ i)\langle\uparrow\rangle$*)
\end{lstlisting}

The \texttt{apply} tactic fails if we try to apply $mstep\_inst$ because the renaming operation does not match with the instantiation operation, meaning that originally, the renaming operation has to be manually changed to an instantiation. 
We apply $mstep\_inst$ with the \texttt{as\_apply} tactic.
After pre-processing, the goal has changed to $(\sigma \ i)[\uparrow] \ggcurly (\tau \ i)[\uparrow]$. The $\sigmamu$-function first solves the equation $(\sigma \ i)[\uparrow] = ?s[?\sigma]$ resulting in $\{?s \mapsto \sigma \ i, ?\sigma \mapsto \uparrow\}$, and solves next $(\tau \ i)[\uparrow]=?t[\uparrow]$ resulting in $\{?t \mapsto \tau \ i\}$. \end{example}

\begin{example}
Consider the following case in the context morphism lemma in the POPLMark challenge:
	\begin{lstlisting}[language=Coq, escapeinside={(*}{*)}]
		(*$T\_Tapp$*) : (*$\forall \ \Delta \ \Gamma \ s \ A \ A' \ B$*), (*$\Delta; \Gamma \vdash s: \forall A. B \to$*) (*$\Delta \vdash A' <: A \to$*) (*$\Delta; \Gamma \vdash s \ A' : B[A' \cdot id]$*)
		(*$H_1$*) : (*$s[\sigma;\tau]$*) : (*$(\forall A . B)[\sigma]$*)
		(*$H_2$*) : (*$\Delta \vdash A'[\sigma] : A[\sigma]$*)
		--------------------------------------------------
		(*$\Delta; \Gamma \vdash s[\sigma;\tau] \ A'[\sigma] : B[A'[\sigma] \cdot \sigma]$*)
	\end{lstlisting}

The \texttt{apply} tactic fails when applying $T\_Tapp$, and this is the reason for the originally indirect definition of $T\_Tapp$. We apply $T\_Tapp$ with \asapply. After the pre-processing steps, we have to match $\Delta;\Gamma \vdash s[\sigma;\tau]  \ A'[\sigma]: B[A'[\sigma] \cdot \sigma]$ with $?\Delta; ?\Gamma \vdash ?s \ ?A' : ?B[?A' \cdot id]$. The equation $s[\sigma; \tau] \ A'[\sigma]= ?s \ ?A'$ matches syntactically and $\sigmamu$-function results in $\{?s \mapsto s[\sigma;\tau], ?A' \mapsto A'[\sigma]\}$. But, $B[A'[\sigma] \cdot \sigma] = ?B[A[\sigma] \cdot id]$ can't be matched with $\sigmamu$-rules and hence $\sigmamu$-function fails. However, it has the form $s[t \cdot \sigma]= ?s[t \cdot id]$ described in the last section. Hence, we resolve $?B$ as $B[\Uparrow \sigma]$, and $B[\Uparrow \sigma][A'[\sigma]]$ normalizes to $B[A'[\sigma] \cdot \sigma]$.
\end{example}

\begin{example}
When proving transitivity of subtyping (POPLMark), we have: 
\begin{lstlisting}[language=Coq, escapeinside={(*}{*)}]
	(*$sub\_weak$*) : (*$\forall \Delta \ \Delta' \ \xi \ A_1 \ A_2$*), (*$\Delta \vdash A_1 <: A_2 \to$*) (*$(\forall x. (\Delta \ x)\langle\xi\rangle=\Delta'(\xi \ x))\to$*) (*$\Delta' \vdash A_1\langle \xi \rangle <: A_2\langle\xi\rangle$*)
	(*$H$*) : (*$\Delta \vdash B\langle\uparrow\rangle <: B'\langle\uparrow\rangle$*)
	----------------------------------------------------
	(*$(B\langle\uparrow\rangle \cdot \Delta \circ \uparrow ) \vdash B\langle\uparrow\rangle <: B'\langle \xi \circ \uparrow\rangle$*) 
\end{lstlisting}

We need to apply $sub\_weak$, which is the weakening lemma for subtyping.
Originally, $sub\_weak$ has two versions: one with the above statement, and an auxiliary lemma that has the conclusion $\Delta' \vdash A_1' <: A_2'$ with the extra premises $A_1' = A_1\langle\xi\rangle$ and $A_2' = A_2\langle\xi\rangle$ which is applied in this case.
We proceed with \asapply.
All equations match syntactically except $B'\langle\xi\circ \uparrow\rangle =?A_2\langle\uparrow\rangle$. Since $B'\langle \xi\circ\uparrow\rangle$ can be decomposed as $B'\langle\xi\rangle\langle\uparrow\rangle$, the $\sigmamu$-function resolves $B'\langle\xi\circ \uparrow\rangle=?A_2\langle\uparrow\rangle$ as $\{?A_2 \mapsto B'\langle\xi\rangle\}$ .
\end{example}

\begin{example}
We look at the proof of $ty\_subst$ from the POPLMark challenge:
\begin{lstlisting}[language=Coq, escapeinside={(*}{*)}]
	(*$sub\_subst$*) : (*$\forall \Delta \ \Delta' \ \sigma \ A \ B$*), (*$\Delta \vdash A <: B \to$*) (*$(\forall x. \Delta' \vdash \sigma \ x <: (\Delta \ x)[\sigma])\to$*) (*$\Delta' \vdash A[\sigma] <: B[\sigma]$*)
	(*$H$*) : (*$\Delta \vdash A' <: A$*)
	----------------------------------------------------
    (*$\Delta' \vdash A' <: A$*) 
\end{lstlisting}

Here, we have a goal in uninstantiated form and an assumption in instantiated form. 
Originally, $A$ and $A'$ are manually changed to $A[id]$ and $A'[id]$ before applying $sub\_subst$.
We apply $sub\_subst$ with \asapply. Eventually, we need to match $A'=?A[?\sigma]$ and $A=?B[?\sigma]$. While the $\sigmamu$-function fails in both cases, our additional heuristics
resolve $?A$, $?B$ and $?\sigma$ as $A'$, $A$ and $id$ respectively. 
\end{example}

\paragraph{Discussion.} Even in relatively short proof scripts such as the POPLMark challenge (642 lines) and POPLMark Reloaded (683 lines), the \asapply \ is used frequently: 15 times in the POPLMark B and 10 times in the POPLMark Reloaded. In all cases, it successfully solves the intended goal. 
Compared to the previous solutions, the proposed solutions with the matching tactics avoid the need for indirect definitions, auxiliary lemmas, and the need to manually transform the goal or find the instances for existential variables.

Additional (single) examples, for example, as one in a formalisation of Martin-L\"of Type Theory \cite{adjedj2024martin},  can be found in the appendix. 
These developments require a customised matching tactic due to manual adaptations of the substitution primitives in the original developments.
Apart from these technical problems, both in this case study and in a development of the call-by-push-value \cite{forster2019call}, we have so far not found any essential problems with the tactic.

\section{Related Work}
Mechanizing syntax has a long history, leading to many syntax representations \cite{de1972lambda, aydemir2008engineering, mcbride2004functional, pfenning1988higher, pitts2001nominal, gordon1993mechanisation, polesiuk2024functorial} and supporting tools \cite{schafer2015autosubst, stark2019autosubst, aydemir2010lngen, keuchel2016needle, sewell2010ott, urban2012general, van2025animating, pientka2010beluga, gacek2008abella, schurmann2009twelf}. Autosubst \cite{schafer2015autosubst, stark2019autosubst} provides automation for de Bruijn syntax and has been used in several mechanizations \cite{abel2017poplmark, forster2019call, adjedj2024martin, forster2021completeness}.

Unification in languages with binders has been studied extensively \cite{goubault2001higher}. Traditionally, unification is defined modulo $\beta\eta$-equivalence, known as higher-order unification.
While higher-order unification is undecidable in general, Huet's algorithm works well in practice \cite{huet1975unification}. 
Miller et al. \cite{miller1991logic} identify a fragment named \textit{higher-order patterns} for which higher-order unification is decidable and is unitary, i.e.~admits general unifiers.
Higher-order matching is decidable up to fourth order \cite{padovani2000decidability}
while for higher orders, decidability remains open. 

Calculi for explicit substitutions such as the $\sigma$-calculus were introduced by Abadi et al. \cite{abadi1989explicit} to bridge the gap between the $\lambda$-calculus and its implementations. Curien et al. \cite{curien1996confluence} prove that the $\sigma$-calculus is a convergent rewriting system, and later Sch\"afer et al. \cite{schafer2015} show that the $\sigma$-calculus is a sound and complete model for the de Bruijn algebra. 
Together, this means the $\sigma$-calculus has practical usage as a rewriting system to decide equality in languages with binders as in the Autosubst library \cite{schafer2015autosubst,stark2019autosubst}.
However, these techniques do not directly extend to matching problems.

Dowek et al. \cite{dowek1995higher} show that higher-order unification is reducible to unification in the first-order equational theory of $\lambda\sigma$-calculus ($\sigma$-calculus with $\beta$ rule), and provide a general unification method for the $\lambda\sigma$-calculus.
Moura et al. \cite{de2005second} show that this method does not decide second-order matching in the $\lambda\sigma$-calculus by providing a non-terminating counter-example. They characterise a fragment for which the method terminates and provide a second-order matching algorithm for this fragment. 
This is not directly applicable as it covers the $\sigma$-calculus, including the $\beta$ rule.

Nominal syntax \cite{gabbay2002new} provides an alternative representation of languages with binders. Urban et al. \cite{urban2004nominal} show that \textit{nominal unification} is both decidable and unitary. Cheney  \cite{cheney2005relating} shows that higher-order pattern unification problems can be solved by encoding them as nominal unification problems, and Levy et al. \cite{levy2008nominal} show the reverse. Nantes-Sobrinho et al. \cite{nantes2025nominal} generalise nominal unification problems to \textit{nominal equational problems} and provide a rule-based algorithm to find solutions in the ground nominal algebra \cite{gabbay2002new}.

Rocq's original unification algorithm relies on heuristics \cite{ziliani2015unification}. Ziliani et al. \cite{ziliani2015unification} give a new unification algorithm along with an implementation for Calculus of Inductive Constructions, incorporating canonical structures and universe polymorphism. They formally describe a heuristic called \textit{controlled backtracking} used in the unification algorithm of Rocq.

\section{Conclusion and Ongoing Work}
We extended the Autosubst framework to generate a pattern-matching tactic for custom syntax. The tactic is intended as a substitute for the \texttt{apply} tactic when matching with an assumption up to the substitution calculus.
We tested this tactic on standard benchmarks, such as the POPLMark \cite {aydemir2005mechanized} and POPLMark Reloaded \cite {abel2017poplmark} challenges, and were able to simplify previous solutions. 


The current tactic operates heuristically for pragmatic reasons -- the $\sigma$-matching problem could potentially have many and even infinite solutions.
However, in case the solution is not unique, it could produce an incorrect solution when there is a correct solution.
Eventually, we are interested in having a matching tactic with proven guarantees; 
for example, we are interested in identifying the biggest fragment $\sigma'$ for which the uniqueness criterion holds: If $\theta$ is a solution of $s=_{\sigma'}t$, then any other solution $\theta'$ is $\sigma$-equal to $\theta$. 



\bibliographystyle{eptcs}
\bibliography{generic}

\appendix
\section{More Examples}

We look at more examples from the MLTT mechanization \cite{adjedj2024martin}\footnote{https://github.com/CoqHott/logrel-coq/blob/coq-8.19/theories/LogicalRelation/Neutral.v}, POPLMark Reloaded, and POPLMark to demonstrate the usefulness of our tactic.

\begin{example}
In the MLTT mechanization, in the proof of $complete\_Pi$, we come across the following.

\begin{lstlisting}[language=Coq,escapeinside={(*}{*)}]
(*$ty\_app$*) : (*$\forall$*) (*$\Gamma$*) ((*$f \ d \ dom \ cod:tm$*)), (*$\Gamma  \vdash f: \Pi_{dom} \ cod$*) -> (*$\Gamma \vdash d: dom $*) -> (*$\Gamma \vdash  tApp \ f \ d : cod[d \cdot id]$*)
(*$H$*) : (*$\Gamma \vdash a: dom\langle\xi\rangle$*)
-------------------------------------------
(*$\Gamma \vdash tApp \ n\langle\xi\rangle \ a \ : \ cod[a \cdot \xi]$*)
\end{lstlisting}

Here, we need to apply the type application rule $ty\_app$ for the $\Pi$ type. The \texttt{apply} tactic would fail, and we resort to \texttt{as\_apply}. First, we do the pre-processing steps, and all the equations are matched except $cod[a \cdot \xi] = ?cod[a \cdot id]$. This equation can't be matched in the $\sigmamu$ fragment. But, it's of the form $s[t \cdot \sigma] = ?s[t \cdot id]$. Since $\xi$ is a renaming, we resolve $?cod$ as $cod\langle\Uparrow \xi\rangle$. After, $cod\langle\Uparrow \ \xi \rangle[a \cdot id]$ is reduced to $cod[a \cdot \xi]$.
\end{example}

\begin{example}
This example is a continuation from the above example, where we have to prove the sub-goal (first premise of $ty\_app$) by applying the weakening lemma $ty\_wk$.
\begin{lstlisting}[language=Coq,escapeinside={(*}{*)}]
(*$ty\_wk$*) : (*$\forall$*) (*$\Gamma \ \Delta \ \xi$*), (*$\vdash \Delta$*) -> (*$\Gamma \vdash t : A$*) -> (*$\Delta \vdash t\langle \xi\rangle : A\langle\xi\rangle$*)
-------------------------------------------
(*$\Gamma \vdash n\langle\xi\rangle : \Pi_{dom\langle\xi\rangle} \ cod\langle\Uparrow \xi \rangle $*)
\end{lstlisting}

The \texttt{apply} tactic fails as expected. We apply the \texttt{as\_apply} tactic on $ty\_wk$ and eventually we face the $\sigma$-matching equation $\Pi_{dom\langle\xi\rangle} \ cod\langle\Uparrow \xi\rangle = ?A \langle\xi\rangle$. This equation matches in the $\sigmamu$ fragment because we have the instantiation law for renaming $(\Pi_{dom} \ cod)\langle\xi\rangle = \Pi_{dom\langle\xi\rangle} \ cod\langle\Uparrow \xi\rangle$. Hence, the $\sigmamu$-function resolves $?A \mapsto \Pi_{dom} \ cod $.
\end{example}

\begin{example}
    In the $step\_inst$ lemma from the POPLMark Reloaded, we face the following.
\begin{lstlisting}[language=Coq,escapeinside={(*}{*)}]
(*$\beta$*) : (*$\forall s \ t$*), (*$(\lambda.s) \ t \succ s[t \cdot id]$*) 
----------------------------
(*$\lambda. s[\Uparrow \sigma] \ t[\sigma] \succ s[t[\sigma] \cdot \sigma]$*)
\end{lstlisting}

The \texttt{apply} tactic can syntactically match the reducible expression $(\lambda.s[\Uparrow \ \sigma]) \ t[\sigma]$ with $(\lambda. s) \ t$ in the $\beta$-rule. But \texttt{apply} fails to match $s[t[\sigma] \cdot \sigma]$ with $s[\Uparrow \ \sigma][t[\sigma] \cdot id]$ because it relies on the Rocq unification engine that doesn't convert up to $\sigma$-rules. In the case of \texttt{as\_apply}, we resolve the variables $?s$ and $?t$ as $s[\Uparrow \ \sigma]$ and $t[\sigma]$ respectively. Finally, we do a $\sigma$-normalization that reduces $s[\Uparrow \ \sigma][t[\sigma] \cdot id]$ to $s[t[\sigma] \cdot \sigma]$. 
\end{example}

\begin{example}
We look at the proof of the context morphism lemma from the POPLMark. At one point in the proof, we have the context and goal as follows:
\begin{lstlisting}[language=Coq,escapeinside={(*}{*)}]
(*$context\_renaming\_lemma$*) : (*$\forall \Delta' \ \Delta \ \Gamma \ \Gamma' \ s \ A \ \xi \ \zeta$*) ... -> (*$\Delta'; \Gamma' \vdash s: A$*) -> (*$\Delta; \Gamma \vdash s\langle\xi; \zeta\rangle : A\langle\xi\rangle$*) 
-------------------------------------------------
(*$\Delta' ; A[\sigma] \cdot \Gamma' \ \vdash \ (\tau \ f) \langle id; \uparrow\rangle : (\Gamma \ f)[\sigma]$*)
\end{lstlisting}

The \texttt{apply} tactic fails when we try to apply the context renaming lemma. The \texttt{as\_apply} proceeds as expected. The equation $(\tau \ f)\langle id; \uparrow\rangle = ?s\langle?\xi; ?\zeta\rangle$ matches syntactically and we resolve $\{?s \mapsto \tau \ f, ?\xi \mapsto id, ?\zeta \mapsto \uparrow\}$. Now, the $\sigmamu$-function fails for the equation $(\Gamma \ f)[\sigma]= ?A\langle id\rangle$. But it has the form $s = ?t\langle id\rangle$ (for renamings). Hence, we resolve $?A$ as $(\Gamma \ f)[\sigma]$
\end{example}

\end{document}